\def\kpc{\;{\rm kpc} }

\def\kms{\;{\rm km}{\rm s}^{-1}  }

\def\spose#1{\hbox to 0pt{#1\hss}}
\def\lta{\mathrel{\spose{\lower 3pt\hbox{$\sim$}}
    \raise 2.0pt\hbox{$<$}}}
\def\gta{\mathrel{\spose{\lower 3pt\hbox{$\sim$}}
    \raise 2.0pt\hbox{$>$}}}
\documentclass{aastex}
\usepackage{amsmath}
\usepackage{emulateapj5}
\begin{document}

\title{First Clear Signature of an Extended Dark
Matter Halo in the Draco Dwarf Spheroidal}
\author{Jan T. Kleyna$^1$, Mark I. Wilkinson$^1$, N. Wyn Evans$^2$, 
Gerard Gilmore$^1$}
\affil{$^1$Institute of Astronomy, Madingley Road,Cambridge, CB3 OHA, UK}
\affil{$^2$Theoretical Physics, 1 Keble Road, Oxford, OX1 3NP, UK}

\begin{abstract} 
We present the first clear evidence for an extended dark matter halo
in the Draco dwarf spheroidal galaxy based on a sample of new radial
velocities for 159 giant stars out to large projected radii.  Using a
two parameter family of halo models spanning a range of density
profiles and velocity anisotropies, we are able to rule out (at about
the $2.5\sigma$ confidence level) haloes in which mass follows light.
The data strongly favor models in which the dark matter is
significantly more extended than the visible dwarf. However, haloes
with harmonic cores larger than the light distribution are also
excluded.  When combined with existing measurements of the proper
motion of Draco, our data strongly suggest that Draco has not been
tidally truncated within $\sim 1\kpc$. We also show that the rising
velocity dispersion at large radii represents a serious problem for
modified gravity (MOND).
\end{abstract}


\section{INTRODUCTION}
The central velocity dispersions of many Local Group dwarf spheroidal
(dSph) galaxies are significantly larger than expected for
self-gravitating systems (e.g. Mateo 1998). Assuming virial
equilibrium, the implied $M/L$ ratios reach as high as $\sim 250$,
making the dSphs among the most dark matter dominated systems in the
universe.  Given the apparent absence of dark matter in globular
clusters (e.g. Dirsch \& Richtler 1995), dSphs are also the smallest
dark matter dominated stellar systems in the universe. As such, they
have emerged as crucial testing grounds for competing theories of dark
matter.

Despite their importance, dynamical models of dSphs to date have been
very simple. Most analyses have relied on the use of single mass,
isotropic King models, with their associated assumptions that
mass-follows-light and that the stellar velocity distribution is
isotropic (though see Pryor \& Kormendy 1991 and Lokas 2001 for more
general models). Hitherto, the validity of such assumptions has
remained unchallenged because of the small size of the data sets. When
only small numbers of radial velocities are available, there is a
well-known degeneracy between mass and velocity anisotropy (e.g.,
Binney \& Tremaine 1987).  An increase in the line of sight velocity
dispersion at large radii may by due to either (1) the presence of
large amounts of mass at large radii, or (2) tangential anisotropy in
the velocity distribution. The primary motivation of this {\it Letter}
is to break this degeneracy for Draco by means of improved modelling
and a larger data set with many more stars in the outer parts.

\section{DATA}

Observations were conducted from June 23 to 26 2000 at the 4m William
Herschel Telescope using the AF2/WYFFOS multifibre positioner and
spectrograph.  A total of 284 stars were observed, spanning the
magnitude range $V\approx 17.0$ to $V\approx 19.8$.  Of these, 159
were Draco members (extending to $25^\prime$) with spectra of
sufficient quality to be included in our dynamical analyses. The
median velocity uncertainty for these 159 stars was $1.9\kms$. Of our
stars, 62 were previously observed by Olszewski, Pryor \& Armandroff
(1996): the agreement between our data and the previous data is
consistent with a binary fraction of $\sim 40\%$. Full details of the
data are presented in Kleyna et al. (2001).

Previous studies (e.g., Hargreaves et al. 1996) have reported weak
evidence of net rotational motion in Draco. N-body simulations show
that tidal disruption of dSphs leads to an apparent rotation about an
axis perpendicular to the orbit of the dSph (Oh, Lin \& Aarseth 1995;
Klessen \& Zhao 2001). In our data set, we find that Draco appears to
rotate at $6\kms$ at a radius of $30^\prime$, with a rotation axis
position angle (PA) of 62$^\circ$.  However, a Monte-Carlo analysis
shows that this rotation signal is not statistically significant
(Kleyna et al. 2001). Further, the solar rest-frame proper motion of
Draco (Scholz \& Irwin 1994) implies that the axis for tidally induced
apparent rotation should have a PA of $138^{\circ +23}_{-23}$. The
$3\sigma$ disagreement with the observed rotation axis argues against
Draco having been significantly influenced by tides, a conclusion
supported by the lack of alignment between the PA of Draco's major
axis ($88\pm3^\circ$, Odenkirchen et al. 2001, henceforth OD01) and
its orbit.

From our data set, it is possible to measure the line of sight
velocity dispersion profile for a strongly dark matter dominated dSph
for the first time.  We divide the data into radial bins and use
Bayes' theorem to obtain the probability distribution and
uncertainties of the velocity dispersion $\langle v^2\rangle^{1/2}$ in
each radial bin (see Kleyna et al. 2001 for details).  The top panel
of Figure 1 shows a plot of the radial variation of $\langle
v^2\rangle^{1/2}$. The velocity dispersion is clearly flat or gently
rising with increasing radius. This plot already rules out the
best-fit mass-follows-light King model (OD01), for which the
dispersion falls to zero at $r_{\rm tidal}=49.4^\prime$, and should
have fallen to $\sim 0.59$ times the central dispersion (or $\sim
5\kms$) by $22^\prime$.

\break

\section{MODELLING}

\subsection{Jeans Equations}

The Jeans equations for a spherical stellar system lead directly to a
model-independent mass estimator (see Binney \& Tremaine 1987,
Eq. 4-55 \& 4-56) requiring only knowledge of the velocity anisotropy
and the true radial velocity dispersion $\langle v_r^2
\rangle$. Assuming spherical symmetry, it is straightforward to obtain
$\langle v_r^2 \rangle$ from the line of sight velocity dispersion
$\langle v^2 \rangle$ using Abel integrals. We consider two dispersion
profiles which bracket the true situation: (1) a flat profile with
amplitude $9.3\kms$ (2) a profile rising linearly from $8.5\kms$ in
the centre to $11.25\kms$ at $22^\prime$. The light is assumed to
follow a Plummer law with a core radius $r_0=9.7^\prime$, as this is
an excellent fit to the star count data (Kleyna et al. 2001).  The
solid and dotted curves in the lower panel of Figure~1 show the mass
profiles obtained based on these two dispersion profiles, assuming
velocity isotropy. The mass enclosed within three core radii ranges
from $6.3 - 18.0 \times 10^7$M$_{\odot}$, implying $M/L$ ratios of
$350-1000$, where the V-band luminosity of Draco is $1.8\times
10^5$L$_{\odot,{\rm V}}$ (Irwin \& Hatzidimitriou 1995, henceforth
IH95).

\vfill

\epsscale{0.8}
\begin{center}
\plotone{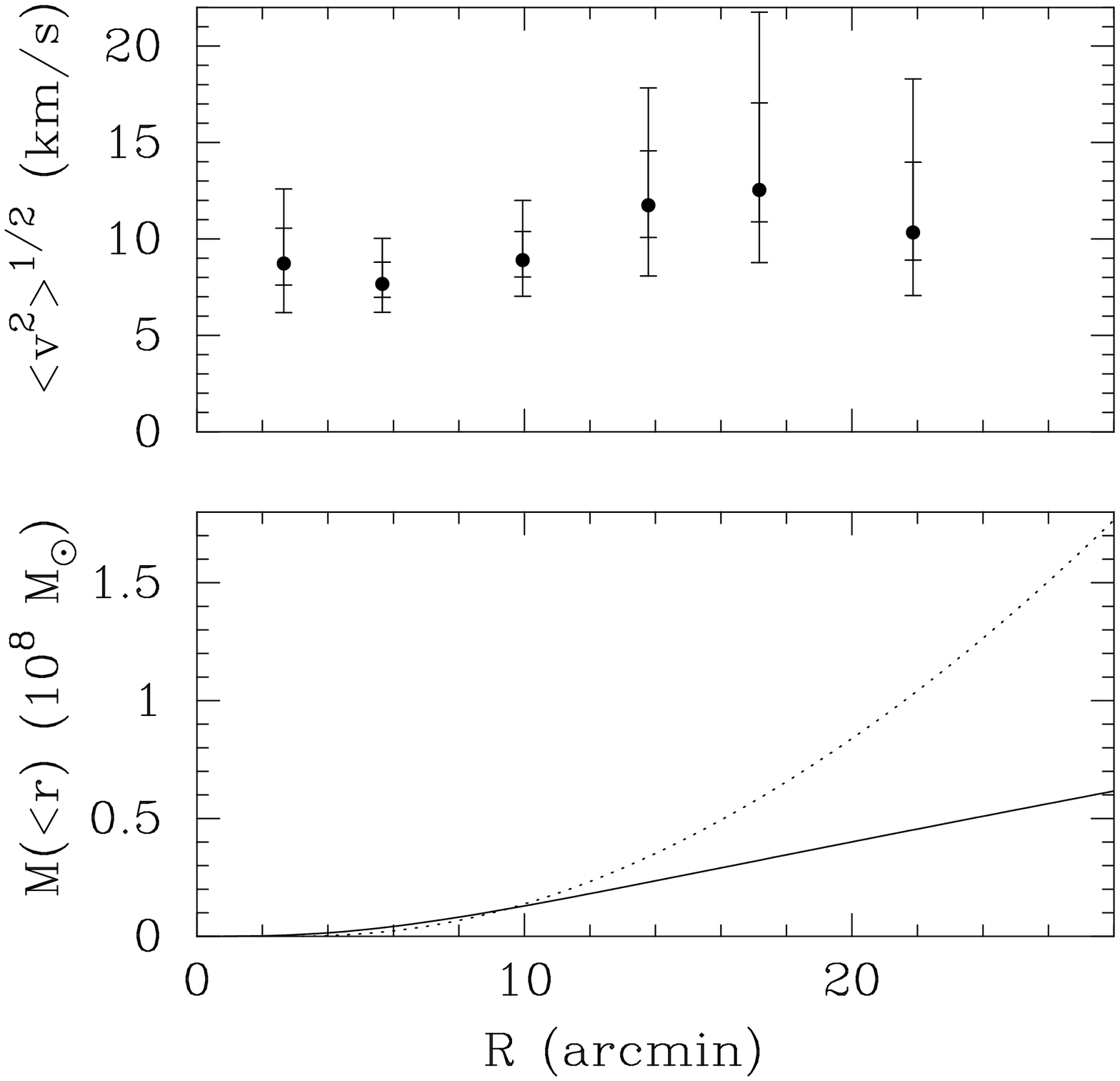} 
\end{center}  
\epsscale{1.0}     
\begin{small}
Fig. 1.-- Top: Line of sight velocity dispersion as a function of
projected radius $R$. Tick marks show the $1$ and $2\sigma$ errorbars
assuming a binary fraction of $40\%$. Bottom: Three-dimensional mass
profiles for Draco obtained using the Jeans equations.
\end{small}

\vfill

\epsscale{0.9}
\begin{center} 
\plotone{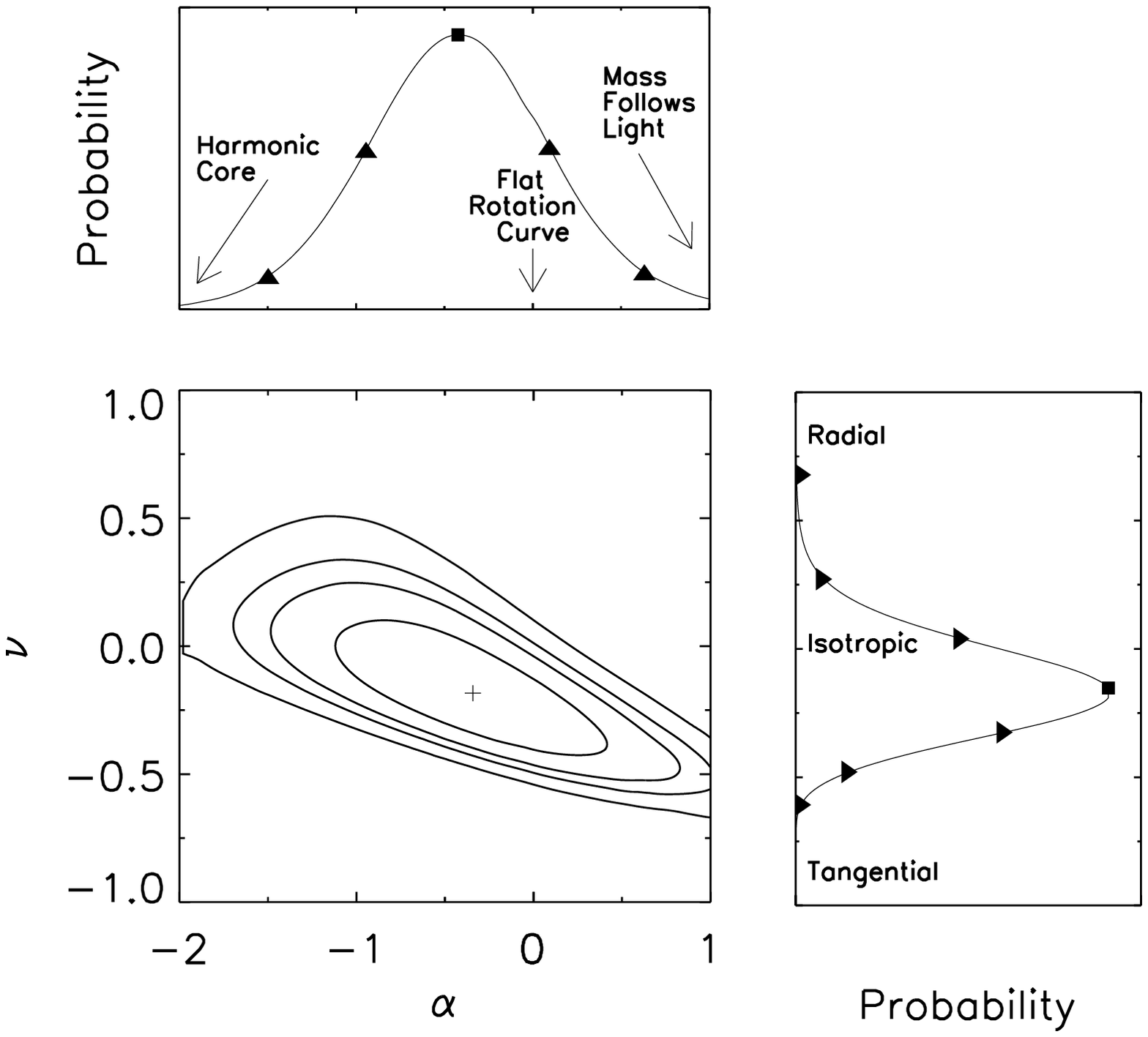}
\end{center}
\epsscale{1.0}
\begin{small}
Fig. 2.--- Likelihood contours of the fit of our Draco data to the
two-parameter $\alpha,\nu$ models of Wilkinson et al. (2001). The
contours are at enclosed two-dimensional $\chi^2$ probabilities of
0.68, 0.90, 0.95, and 0.997. The most likely value is indicated by a
plus sign. The top and right panels represent the probability
distributions of $\alpha$ and $\nu$, respectively; the median of each
distribution is represented by a square, and the triangles show the
$1\sigma$, $2\sigma$ and $3\sigma$ limits. The $3\sigma$ limits are
omitted from the projection of $\alpha$ as the upper limit is strongly
determined by the cut-off at $\alpha = 1$. A $40\%$ binary fraction is
assumed.
\end{small}
\\

\subsection{Halo Models and Distribution Functions}

We now use a two parameter family of models which spans a range of
possibilities for the halo of Draco. The halo potential $\psi(r)$ is
given by
\begin{equation}
\psi(r) = 
\begin{cases}   
\psi_0 (1+r^2/r_0^2)^{-\alpha/2} & \text{if $\alpha \neq 0$},\\
(v_0^2/2)\log\left[1+ r^2/r_0^2\right]& \text{if $\alpha =0$}.
\end{cases}
\label{eq:pot}
\end{equation}
The parameter $\alpha$ determines the mass distribution of Draco's
halo. If $\alpha = 1$, the dark matter has the same density
distribution as the light. In models with $\alpha = 0$, the halo has
an asymptotically flat rotation curve. Finally, when $\alpha=-2$, the
visible dwarf lies in the approximately harmonic core of a much larger
halo.  The distribution functions for dSph density profiles in
potentials (\ref{eq:pot}) are derived in Wilkinson et al. (2001). They
contain a second parameter $\nu$ which measures the velocity
anisotropy.  Isotropic models have $\nu = 0$ and models with radial
(tangential) anisotropy have $\nu > 0$ ($\nu < 0$).  We generate line
profiles from our models by numerically integrating the distribution
functions over all transverse velocities. The line profile is then
convolved with a binary distribution and Gaussian velocity measurement
uncertainty profile of width $\sigma$ to give $\tilde
L(v,R,\sigma;\alpha,\nu)$. The probability of observing an ensemble of
points $R_i$ with velocity $v_i$ and velocity uncertainty $\sigma_i$
is then given by the product of the $\tilde
L(v_i,R_i,\sigma_i;\alpha,\nu)$ values.

Figure~2 shows the likelihood contours for the model parameters
$\alpha,\nu$ assuming a $40\%$ binary fraction. The most likely values
are $\alpha = -0.34$ and $\nu = -0.18$, indicating a model with
significant dark matter at large radii and a tangentially anisotropic
velocity distribution. We rule out both a mass-follows-light ($\alpha
= 1$) and a harmonic core ($\alpha = -2$) model at the $\sim
2.5\sigma$ confidence level.  Our best fit model implies a mass inside
3 core radii of $8^{+3}_{-2} \times 10^7$M$_{\odot}$. The implied
V-band $M/L$ ratio of Draco (within 3 core radii, or $29^\prime$) is
$\sim 440 \pm 240$. This mass profile lies between the two curves in
Figure~1.

All our halo models (\ref{eq:pot}) have cores.  Cosmological
simulations favor cusped halos, such as the Navarro-Frenk-White (NFW)
profile.  The present data are insufficient to discriminate strongly
between cored and cusped profiles for Draco's dark halo -- unlike the
case of the Milky Way galaxy (Binney \& Evans 2001). Cosmological
implications of data sets of dSph radial velocities are discussed
elsewhere (Wilkinson et al., in prep.).

\subsection{Perigalacticon and Tidal Cut-off}
\epsscale{0.9}
\begin{center} 
\plotone{f3.eps}
\end{center}
\epsscale{1.0}
\begin{small}
  Fig. 3.--- Perigalacticon of Draco's orbit versus orbital
  eccentricity. Contours are $1\sigma$, $2\sigma$ and $3\sigma$
  confidence levels. Overplotted solid curves show the relationship
  between perigalacticon and eccentricity assuming the labelled values
  of the tidal radii and our best-fit model for the halo of Draco. The
  dotted curve shows the earlier King model fit to Draco. The solid
  curves in the top and right panels are the probabilities of $e$ and
  $D_p$, respectively, and the dashed curve in the right panel shows
  the distribution of $D_p$ assuming no proper motion information.
\end{small}
\\

\subsection{Perigalacticon and Tidal Cut-off}
A dSph in orbit about the Milky Way is truncated by the Galactic tidal
field during perigalacticon passages (Oh \& Lin 1992). Based on SDSS
images of Draco, OD01 find no evidence for a tidal tail of Draco
beyond $50^\prime$ and no break in its luminosity profile. Thus it
seems likely that the actual tidal truncation radius of Draco is $\gta
1^\circ$.

We use the radial velocity and proper motions to generate possible
Draco orbits. The proper motions are given by Scholz \& Irwin (1994)
as: $\mu_{\alpha}\cos\delta = 0.09\pm0.05''$ cent$^{-1}$,
$\mu_{\delta} = 0.1\pm0.05''$ cent$^{-1}$. The line of sight velocity
(corrected for the solar motion) is $-98\pm 1\kms$, while the distance
of Draco is $82 \pm 7$ kpc (Mateo 1998). We represent the Galactic
potential by $\psi(r) = v_0^2 \log[(\sqrt{a^2+r^2} + a)/r]$ (e.g.,
Wilkinson \& Evans 1999) where the scale length $a \sim 170$kpc and
the normalization $v_0$ is obtained by taking the circular speed to be
$220\kms$ at the solar radius. We assume log-normal errors on $a$ with
$3\sigma$ range $44$kpc to $489$kpc; these limits correspond to masses
within $100$kpc of $4.6\times 10^{11}$M$_\odot$ and
$1.1\times10^{12}$M$_\odot$, respectively. Assuming Gaussian errors on
all other observed quantities, we generate sets of initial conditions
consisting of a position and velocity for Draco and a scale length for
the Milky Way.  Each set is integrated in the Galactic potential for a
few orbital periods and the perigalacticon $D_{\rm p}$ and
eccentricity $e$ of the orbit is determined. Dynamical friction is
unimportant on such short time scales.  The contours in Figure~3 show
the $1$, $2$ and $3\sigma$ limits on the perigalacticon and
eccentricity of Draco's orbit. While the distribution of
eccentricities is roughly flat in the range $[0.3,0.95]$, the
distribution of perigalacticons is strongly peaked at $\sim 75$
kpc. This suggests that Draco is presently at or close to
perigalacticon.

The tidal radius for a dSph moving in the above Galactic potential is
given by (c.f. King 1962)
\begin{equation}
r_{\rm t}^3 = \frac{M_{\rm D}(r_{\rm t})D_{\rm p}^3}{M_{\rm
g}(1-e)}\left[\frac{(1+e)^2}{2 e
v_0^2}\left(\psi_{\rm p} - \psi_{\rm a}\right) + \frac{a(a^2+2 D_{\rm
p}^2)}{(a^2+D_{\rm p}^2)^{3/2}}\right]^{-1},
\end{equation}
where $M_{\rm D}(r_{\rm t})$ is the mass of Draco within radius $r_t$,
$M_{\rm g}$ is the total mass of the Galaxy, and $\psi_{\rm p}$ and
$\psi_{\rm a}$ are the values of the potential at perigalacticon and
apogalacticon. The solid curves in Figure~3 show the variation of
$D_{\rm p}$ with $e$ assuming a range of values for $r_{\rm t}$. Only
models with $r_t > 3^{\circ}$ fall within the $1\sigma$ contour of
Draco's perigalacticon and eccentricity derived from its space
velocity.

If Draco had recently been subject to strong tidal forces, we would
expect significant elongation of the stellar distribution, in conflict
with the roundish ($\epsilon = 0.3$) appearance of Draco on the sky
and the absence of tidal tails (OD01).  Chance alignment effects (e.g.
Kroupa 1997) are unlikely to be responsible for Draco's regular
appearance. Since any tidal extension of Draco should be aligned with
its orbit, elongation approximately along the line of sight would
require Draco's orbit to be similarly aligned. However, Monte Carlo
sampling of the orbital velocities using the observational errorbars
rules out alignment within $45^\circ$ with $98\%$ confidence, assuming
that the prior (no proper motion information) probability of the line
of sight angle is uniform. The absence of statistically significant
apparent rotation argues further against line of sight elongation.

\section{MOND AND DRACO}
As an alternative to dark matter, Milgrom (1983) proposed a
modification to Newton's law of gravity (MOND) at low accelerations. A
number of attempts have been made to apply MOND to the dSphs of the
Milky Way (Gerhard 1994, Milgrom 1995). These used only the global
velocity dispersion to obtain a MOND $M/L$ ratio and were unable to
construct an unambiguous case for or against MOND. Lokas (2001), using
published velocities, obtained MOND fits to the dispersion profiles of
several dSphs, treating the global MOND acceleration scale as a free
parameter.

We model the luminosity density $I(r)$ of Draco using a Plummer sphere
with core radius $r_0$. The velocity anisotropy parameter $\beta$
(Binney 1981) is assumed to vary with radius as $\beta(r) = (\gamma/2)
r^2/(r_0^2+r^2)$ yielding a velocity distribution which approaches
isotropy in the centre and becomes increasingly radial (tangential) if
$\gamma$ is positive (negative). With these assumptions, the Jeans
equations can be integrated to obtain the radial velocity dispersion
$\sigma_r^2(r)$ as
\begin{equation}
\sigma_r^2(r) = - \frac{1}{I(r)(r_0^2+r^2)^{\gamma/2}}
\int_r^{\infty}(r_0^2+r^2)^{\gamma/2} I(r) g_{\rm M}(r)\, {\rm d}r,
\end{equation}
where $g_{\rm M}(r)$ is the MOND acceleration due to the mass interior
to radius $r$.  The central projected V-band luminosity density $I_0$
is $2.2\times 10^6$L$_{\odot}\kpc^{-2}$ (IH95).  Since all the mass in
the MOND picture is provided by the stars, the Newtonian acceleration
$g_{\rm N}(r)$ is given simply by $g_{\rm N}(r)=-G M_{*}(r)/r^2$,
where the stellar mass $M_{*}(r)$ is obtained from $I(r)$ and depends
on the $M/L$ ratio of the stars. The MOND acceleration is then obtained
via $g_{\rm M} = g_{\rm N}/\mu(g_{\rm M}/a_0)$, where $a_0 = 2 \times
10^{-8}$ cm\,s$^{-2}$ is the MOND acceleration scale, $\mu(x)
\rightarrow 1$ for $x \gg 1$ and $\mu(x) \rightarrow x$ for $x \rightarrow 0$. 
Our assumed $a_0$ is an upper limit to the likely value (Sanders \&
Verheijen 1998), and favors models with lower stellar $M/L$ ratios.

Our analysis assumes that Draco is an isolated MOND stellar
system. This is justified by noting that for stars between $\sim
1.7^\prime$ and $\sim 20^\prime$ the acceleration due to the baryonic
mass of the Milky Way disk ($\sim 6\times 10^{10}$M$_{\odot}$) at the
distance of Draco is smaller than the internal accelerations. Since
only $\sim 10\%$ of our sample lies outside $20^\prime$ the assumption
of isolation is reasonable. Representing a slightly flattened light
distribution by a spherical model has greater impact when applying
MOND since all the mass is in the stars. However, in the case of Draco
this effect is not significant as our spherical model underestimates
the mass interior to radius $r$ by $\lesssim20\%$, resulting in $M/L$
ratios which are at most $20\%$ overestimated.

From the projected dispersions, we obtain the MOND line profiles
(assumed Gaussian) which we convolve with the binary velocity
distribution and the error distribution to obtain the observable line
profiles. Using these profiles, we analyse the velocity data set as in
Section 3.2, except that the model parameters are now the stellar $M/L$
ratio and the velocity anisotropy. Figure 4 shows the contours
in the space of $M/L$ and $\nu$. Again, we find that some
tangential anisotropy is required to reproduce the flat projected
velocity dispersion profile. More interestingly, the most likely $M/L$
ratio is 19 with a $3\sigma$ lower limit of 8.6. The assumed value of
$I_0$ implies a total luminosity of $3.2\times 10^5$L$_{\odot}$ out to
$3r_0$, approximately the $2\sigma$ upper limit of the V-band
luminosity (IH95). Our lower bound is therefore a true lower limit on
the MOND $M/L$ ratio of Draco.

Estimates of the $M/L$ ratio of Draco's stellar population based on
comparison with globular clusters have a $3\sigma$ range of $1-4.3$
(Parmentier \& Gilmore 2001, Feltzing, Gilmore \& Wyse 1999). Even our
$3\sigma$ lower limit on the MOND $M/L$ ratio of Draco is at variance
with these values. Thus, even by modifying the law of gravity as
prescribed by MOND, we are unable to explain the observed motions of
Draco's stars without the presence of significant quantities of dark
matter.

\epsscale{0.9}
\begin{center} 
\plotone{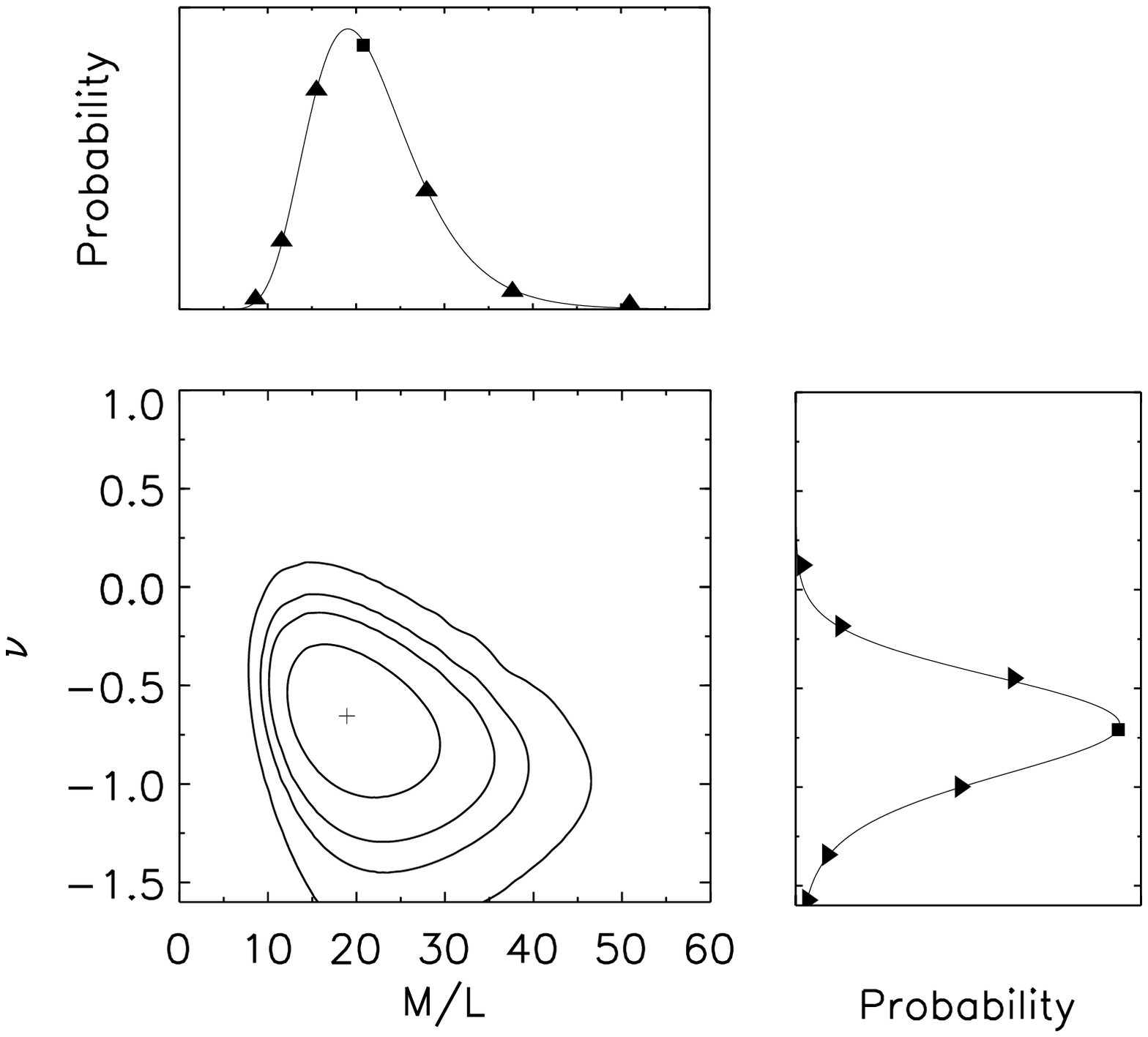}
\end{center}
\epsscale{1.0}
\begin{small}
  Fig. 4.--- Likelihood contours for Draco's $M/L$ and the anisotropy
  parameter $\nu$ assuming the validity of MOND.  The contours are at
  enclosed two-dimensional $\chi^2$ probabilities of 0.68, 0.90, 0.95,
  and 0.997. The most likely mass to light ratio is $M/L=19
  M_\odot/L_\odot$, and $M/L=8.6 M_\odot/L_\odot$ is ruled out at the
  $3\sigma$ level.
\end{small}
\\


\section{DISCUSSION AND CONCLUSIONS}

This {\it Letter} has presented a new set of radial velocities for 159
stars in the Draco dSph.  Our data extend to $\sim 25^\prime$ and are
the first observations to probe the outermost regions of a strongly
dark matter dominated dSph.  The velocity dispersion profile is flat
or slowly rising at large radii, which provides the first clear
signature of an extended dark matter halo in any dSph. We analyse
these data using both the Jeans equation and halo models with
distribution functions.  Our best fit model has a total mass within $3
r_0 \approx 29^\prime$ of $8^{+3}_{-2}\times 10^7$M$_{\odot}$ and a slightly
tangentially anisotropic velocity distribution. We are able to rule
out the traditional mass-follows-light models and extended harmonic
core models at about the 2.5$\sigma$ significance level.

From orbit integrations in a Galactic potential, we argue that Draco
is currently close to perigalacticon and its actual tidal radius is
$> 1^\circ$, several times the characteristic length scale of the
luminous matter. Draco's lack of significant apparent rotation and the
significant misalignment of its major axis with its orbit argue
against tidal forces having played a major role in its evolution. Our
results also have consequences for the Modified Newton Dynamics
(MOND): a Bayesian analysis of the observational data, incorporating
radially varying velocity anisotropy, yields a $3\sigma$ lower limit
of 8.6 on the MOND $M/L$ ratio still requiring the presence of a
significant amount of dark matter in Draco.

\acknowledgments
NWE is supported by the Royal Society, while MIW and JK acknowledge
help from PPARC. The authors gratefully thank Colin Frayn for help
during the observations, and Mike Irwin, HongSheng Zhao, and
the referee Slawomir Piatek for valuable discussions.

\eject

\end{document}